\documentstyle[11pt,twoside,jltp]{article}

\title{Tunneling into 1D and quasi-1D conductors and  Luttinger-liquid
behavior}

\author{E.B. Sonin\address{The Racah Institute of Physics, The Hebrew 
University of Jerusalem,
Jerusalem 91904, Israel \\ and \\ Low Temperature Laboratory, Helsinki
University of  Technology,
FIN-02015 HUT, Finland}}

\runninghead{E.B. Sonin}{Tunnelling into 1D and quasi-1D conductors}
       
\begin{document}

\begin{abstract}
The paper addresses the problem whether and how it is possible to detect the
Luttinger-liquid behavior from the $IV$ curves for tunneling to 1D or
quasi-1D conductors. The power-law non-ohmic $IV$ curve, which is usually
considered as a manifestation of the Luttinger-liquid behavior, can be
also deduced from the theory of the Coulomb blockaded junction between 3D
conductors affected by the environment effect. In both approaches the
power-law exponents are determined by the ratio of the impedance of an
effective electric circuit to the quantum resistance. Though two approaches
predict different power-law exponents (because of a different choice of
effective circuits), the difference becomes negligible for a large number of
conductance channels.

\end{abstract}

\maketitle


\section{Introduction}

One-dimensional and quasi-one-dimensional electron systems are
attracting attention of theorists and experimentalists many years. From
theorists' point of view they are interesting since in a number of
cases one can obtain exact solutions for them taking into account
many-body interactions without using the perturbation theory.
Theoretically it has been well established that in a 1D elctron gas with
{\em arbitrarily} weak interaction Landau's Fermi liquid (FL) theory
breaks down, and the system is expected to behave as a {\em Luttinger
liquid} (LL). The most important feature of the Luttinger liquid, in
contrast to FL, is an absence of the fermion quasiparticle branch at low
energies: excited states of the system must be described by the boson
excitations which correspond to many-body electron states with a huge
number of the   electron-hole pairs (see, e.g., a recent review \cite{FG}
and references therein). This should have a pronounced effect on the
tunneling into a LL conductor: the $IV$ curve of a tunnel junction
between a normal FL and a LL conductor is expected to be non-ohmic
described by a power law with an exponent depending on interaction
strength. 

Experimental evidences for LL behavior have been reported in quantum
wires \cite{QW}, and single walled \cite{Bockrath} and multiwalled
\cite{Schonenberger99} carbon nanotubes. However, it is not completely
clear, to what extent these experimental evidences may be considered as an
unambiguous proof of the LL behavior. Measurements of $IV$ curves for single
walled \cite{Bockrath} and multiwalled \cite{NG} carbon nanotubes  have
revealed a deviation from the LL behavior at high voltages. In Ref.
\cite{NG} they observed a crossover from a non-ohmic $IV$ curve to Ohm's
law with a Coulomb-blockade offset determined by the capacitance of the
junction between  a 3D metallic lead and a nanotube. This crossover is
predicted by the theory of the Coulomb blockade in a junction between normal
FL conductors \cite{SZ,IN}. A crucial role in this theory belongs to the
environment effect, which depends on the ratio of the real part of the
circuit impedance to the quantum resistance $R_K=h/e^2 \approx 26$
k$\Omega$. The environment effect originates from the Nyquist-Johnson noise
in an electric circuit. The noise results in fluctuations of the phase at
the junction, which can suppress the Coulomb blockade. Further we call this
approach the environment quantum fluctuation (EQF) theory. 

The existing LL theory for tunneling takes into account accurately the
Coulomb interaction inside the LL liquid conductor, but ignores the
Coulomb energy of the junction charge, which is defined by the junction
capacitance $C_T$. The fact that the junction Coulomb energy $e^2/C_T$
important not only at low voltages, as widely known \cite{Bockrath}, but
also at high voltages, means that the high-voltage part of the
tunnel-junction $IV$ is a bad probe of bulk properties of nanotubes,
and one should look for evidences of LL behavior mostly at intermediate
voltages characterized by the power-law $IV$ curve. However, both
EQF and LL pictures predict a power-law dependence, and a serious problem
arises, how, and whether it is possible at all, to discriminate these two
pictures. This is the problem that the present paper addresses.

Section \ref{env} reminds the basic ideas of the EQF theory and defines the
power-law exponents for the tunnel conductance and the function $P(E)$,
which determines a probability of the transfer of the energy $E$ to the
environment. In Sec. \ref{Lut-1} the results of the LL theory are
discussed and compared with predictions of the EQF theory for an one-channel
1D conductor with spinless electrons. Section \ref{Lut-2} considers the
same questions, but for a quasi-1D conductor with many channels. The
results of the analysis are summarized in the concluding Sec. \ref{Con}.

\section{The Coulomb blockade and the environment quantum fluctuation
(EQF) theory} \label{env}

We restrict our discussions by zero temperature.
The standard tunneling theory based on the tunneling hamiltonian and the
quantum-mechanical Golden Rule gives the following expression for the
current through the junction:
\begin{equation}
I= {1\over eR_T}\int _0^{eV} dE_1 \int _0^{eV}\rho(E_2) dE_2
\delta(E_1-E_2)~.
\label{IVst}   \end{equation}
Here $\rho(E)$ is the relative density of the
state for the right conductor normalized to the constant DOS of the
normal Fermi-liquid, the latter being included into the definition  of
the junction conductance $1/R_T$. The left conductor is supposed to be
always a FL conductor and the relative DOS for him is unity. If the
right conductor is also a FL conductor then $\rho(E)=1$, and the $IV$
curve is exactly ohmic.

The EQF theory \cite{IN} assunes that both conductors, which form a
junction, are 3D FL conductors, {\em i.e.}, $\rho(E)=1$, but takes into
account  fluctuations in the electric circuit (the Nyquist-Johnson noise).
This results in the fluctuation of the phase of the tunnelling hamiltonian.
Therefore the delta-function in Eq. (\ref{IVst}) should be replaced by the
function which is a Fourier transform of the phase correlator:
\begin{equation}
P(E)={1\over 2\pi \hbar}\int_{-\infty}^\infty dt e^{iEt/\hbar}
\langle e^{i\hat\varphi(t)} e^{-i \hat \varphi(0)}\rangle~,
  \label{P-E}   \end{equation}
and Eq. (\ref{IVst}) becomes
\begin{equation}
I= {1\over eR_T}\int _0^{eV}dE_1 \int _0^{eV} dE_2
P(E_1-E_2)= {1\over eR_T}\int _0^{eV}dE (eV-E)P(E)~.
   \label{IV-P}   \end{equation}
 
The phase  $\varphi$ is connected with the voltage $V$ across the
junction by the Josephson relation
\begin{equation}
\hbar {d\varphi \over dt}=eV~.
  \label{Jos}   \end{equation}
Since the phase $\hat \varphi$ is an  operator conjugate to
the operator of the  electron number, the operator $e^{-i\hat\varphi(t)}$
is a creation operator of the electron in the circuit at the moment
$t$; or, more exactly, since  $ \varphi$ is the phase difference
across the junction, an operator of the electron 
annihilation in the left conductor and electron creation in the right
conductor. The correlator in Eq. (\ref{P-E}) may be presented as
\begin{equation}
\langle e^{i\hat\varphi(t)} e^{i \hat \varphi(0)}\rangle =e^{J(t)}~,
  \label{P-E-J}   \end{equation}
where at $T=0$ the correlator $J(t)$ is
\begin{equation}
J(t)=  \langle[\hat\varphi(t)-\hat\varphi(0)]\hat \varphi(0)\rangle
=2\int_0^\infty {d\omega \over \omega}{\mbox{Re} Z(\omega) \over
R_K}\left(e^{-i\omega t}-1 \right)~.
  \label{J-t}   \end{equation}
This relation is a direct consequence of the fluctuation-dissipation
theorem, which connects the correlator of a ``coordinate'' with a linear
response to a conjugate ``force''. A conjugate ``force'' for the
``coordinate'' $\varphi$ is $\hbar I /e$, and because of the Josephson
relation (\ref{Jos}), the imaginary part of the response is proportional
to the real part of the impedance $Z(\omega)=V(\omega)/I(\omega)$:
$\mbox{Im}\{e\varphi(\omega) /\hbar I(\omega)\}=\mbox{Re}\{e^2
Z(\omega)/\hbar^2\omega \}$. 

The function $P(E)$ gives  probability of the excitation of the
environment mode after the electron tunneling. In the case of
low-impedance environment ($\mbox{Re} Z/R_K \rightarrow 0$) this
probability is neglible and $P(E)$ function reduces to the delta-function
$\delta(E)$. This means that the environment fluctuations eliminate the
Coulomb blockade. In the opposite limit of the high-impedance
environment ($\mbox{Re} Z/R_K\rightarrow \infty$) the function  $P(E)$
becomes $P(E)=\delta(E-E_c)$, where $E_c=e^2/2C_T$ is the Coulomb energy of
the junction. The $IV$ curve of the junction in this case has a voltage
offset $e/2C_T$, which is a manifestation of the Coulomb blockade. For
finite $\mbox{Re} Z/R_K$ at $E$ much smaller than the Coulomb energy, $P(E)$
is a power-law function:
\begin{equation}
P(E) \propto {\tau  \over \hbar} \left(E \tau  \over
\hbar\right) ^{\alpha_E-1}~,
  \label{P-sV}   \end{equation}
where
\begin{equation}
\alpha_E={2\mbox{Re} Z(0)\over R_K}~, 
  \label{alpha}   \end{equation}
and $\tau = \mbox{Re} Z(0) C_T$ is the relaxation time in the effective
electric circuit. 

At zero temperature the probability $P(E)$ determines the second
derivative of the
$IV$ curve:
\begin{equation}
{d^2I\over dV^2} = {e\over R_T}P(eV)~.
   \label{IV-d}   \end{equation}
Thus the exponent $\beta_E$ of the conductance $dI/dV \propto
V^{\beta_E}$ is equal to $\alpha_E$.

In the limit of $E\gg E_c$, $P(E)$ is a decreasing
function of $E$ :
\begin{equation}
P(E)={2\over E}{\mbox{Re} Z(E/\hbar)\over R_K}~.
  \label{P-hV}   \end{equation}

Phase fluctuations at the junction are determined by the impedance of the
effective electric circuit, which includes the capacitance of the junction
\cite{IN}, i.e.
\begin{equation}
{1\over Z(\omega)}= -i\omega C_T + {1\over Z_0(\omega)}~,
     \label{Zh}   \end{equation}
where $Z_0(\omega)$ is the impedance of the circuit connected to the
junction. For a pure ohmic environment, $Z_0=R$, the exponent
$\alpha_E=2R/R_K$ determines also the $IV$ curve and high voltages $V \gg
e/C_T,~\hbar/e\tau$. High voltages probe the high-frequency impedance
$Z(\omega)$. At the relevant high frequency $\omega=eV/\hbar$, the junction
capacitance $C_T$ short circuits the impedance, and the $IV$ curve is
obtained using Eqs. (\ref{IV-P}), (\ref{P-hV}), and (\ref{Zh}):
\begin{equation}
I \approx {1\over R_T}\left[V -{e\over 2C_T} +{\alpha_E \over 2\pi^2}
\left({e\over 2C_T}\right)^2 {1\over V}\right]~. 
  \label{IVh}   \end{equation}
This high-voltage asymptotics, characterized by the Coulomb offset $e/2C_T$
and the ``tail'' voltage $\propto 1/V$  was experimentally studied and
discussed  by Wahlgren {\it et al.}, and Penttil\"a {\it et al.}  within
the horizon picture \cite{WDH,Penttila00}. 

For comparison with the LL theory important is the
case of an infinite transmission $LC$ line with the real impedance  $Z_0
=\sqrt{l_l/c_l}$, where $l_l$ and $c_l$ are the line inductance and the
line capacitance per unit length. Then $\alpha_E=2Z_0/R_K$. A conductor
can be treated as an infinite $LC$ line, if its total ohmic resistance is
much higher then the impedance $\sqrt{l_l/c_l}$, but does not exceed,
nevertheless, its total inductive resistance $\omega l_l {\cal L}$, where
${\cal L}$ is the transmission line length.

\section{Tunneling into an one-channel 1D conductor (spinless electrons)}
\label{Lut-1}
\subsection{Tunneling in the Luttinger-liquid (LL) theory}

For the sake of simplicity we start from the one-channel case of
spinless electrons. In the LL theory the environment effects are absent,
and we must return to Eq. (\ref{IVst}), which after one integration is:
\begin{equation}
I= {1\over eR_T}\int _0^{eV} dE\rho(E) dE~.
      \label{IV-LL}   \end{equation}
Because of the electron-electron interaction the relative DOS 
$\rho(E)$ essentially  different from unity. It is determined by the 
Fourier transform of the electron Green function $\langle \hat \psi(x,t)
\psi^\dagger (x,0)\rangle$, which is determined by the boson degrees of
freedom. For the end contact $\rho(E)$ is given by the relation very similar
to Eq. (\ref{P-sV}) \cite{FG}:
\begin{equation}
\rho(E) \propto  \left(E\tau_c \over
\hbar\right)^{\alpha_L -1}~,
      \label{rho-g}   \end{equation}
where 
\begin{equation}
\alpha_L={ v_{pl}\over v_F} 
  \label{alphaLL}   \end{equation}
is the interaction parameter for the one-channel Luttinger liquid (spinless
fermions), $v_F$ is the Fermi velocity, and 
\begin{equation} 
v_{pl}= \sqrt{v_F^2 +{2 e^2 v_F\over \pi \hbar} \ln{r_g\over r_0}}
      \label{specC}\end{equation}
is the velocity of the boson collective mode in the Luttinger liquid
with the long-range Coulomb interaction. Here $v_F=\pi \hbar n_1/m^*$ is
the Fermi velocity, $n_1$ is the 1D electron density per unit wire length,
$m^*$ is the effective electron mass, $r_0$ is the radius of the wire and
$r_g$ is the distance from the metallic ground. The boson mode is nothing
else but an 1D plasmon mode $\omega = v_{pl} k$. For small $\alpha_L -1$
the correction to the constant Fermi-liquid DOS is small: $\rho(E) \approx
1 +(\alpha_L -1)\ln(E\tau_c/\hbar)$. This correction can be obtained from
the theory of Altshuler and Aronov \cite{AA} using the perturbation
theory. 

The DOS  $\rho(E)$ determines the {\em first} derivative of
the $IV$ cirve (conductance):
\begin{equation}
{dI\over dV} = {1\over R_T}\rho(eV)~.
   \label{IV-dL}   \end{equation}
Thus the exponent of the conductance $\propto V^{\beta_L}$ is
$\beta_L=\alpha_L-1$.
 
Fisher and Glazman \cite{FG} chose the cut-off time
$\tau_c$ in Eq. (\ref{rho-g}) to be of order of the inverse Fermi
energy. Another possible cut-off is related to the wave vector $k_c$ at
which the Coulomb interaction becomes weak \cite{Sch}. According to
Eq. (\ref{specC}) it is the inverse size of the 1D wire (nanotube): $k_c
\sim 1/r_0$. Then the cut-off time in Eq. (\ref{rho-g}) is $\tau_c
\sim r_0/v_F$. This cut-off is more relevant if the wire radius
$r_0$ is essentially larger then the inverse Fermi wave vector. 

\subsection{Comparison the EQF and LL theories}

Let us discuss similarity between the DOS $\rho(E)$ in the LL
theory and the function $P(E)$ in the EQF theory. The former is 
defined as a Fourier transform of the averaged operator product
$\langle \hat \psi(x,t)  \hat \psi^\dagger(x,0)\rangle$ \cite{FG}, and
the latter is a Fourier transform of the correlator $\langle
e^{i\hat\varphi(t)} e^{-i
\hat \varphi(0)}\rangle$. Both the operators $e^{-i\hat\varphi(t)}$ and
$\hat \psi^\dagger(x,t) $ are creation operators of a single charge
$e$. One should expect then similar functional dependencies for $\rho(E)$
and  $P(E)$. This takes place when the voltage at the
junction is less than the Coulomb gap $e/C_T$.  

Indeed, the exponents $\alpha_E$ and $\alpha_L$ for the power laws
predicted by the EQF theory, Eq.  (\ref{alpha}),  and by the LL theory,
Eq. (\ref{alphaLL}), coincide, if one interprets the 1D plasmon mode as
a  wave along the lossless $LC$ transmission line formed by a 1D conductor
and a metallic ground. The capacitance per unit length of the transmission
line is $c_l=1/(2 \ln{r_g/r_0})$, and the inductance per unit length is
determined by the kinetic energy of electrons, {\em i.e.,} is a kinetic
inductance $l_l=m^*/( e^2n_1)=R_K/2v_F$, which essentially exceeds the
geometric inductance  $\ln(r_g/r_0)/ c^2$ because of low electron 1D
density $n_1$  in 1D wires, compared to the density $nS$ in 3D wires with
3D density $n$ and the cross-section area $S$. The transmission line with
these parameters supports the sound-like  wave with the velocity
$1/\sqrt{l_lc_l}$ which coincides with the plasmon velocity $v_{pl}$
given by Eq. (\ref{specC})  in the limit of very strong Coulomb
interaction $v_{pl}\gg v_F$. Using the
impedance $Z=\sqrt{l_l/c_l}=l_l v_{pl}$ of the infinite $LC$ transmission
line, Eq. (\ref{alpha}) for $\alpha_E$ becomes identical to Eq.
(\ref{alphaLL}) for
$\alpha_L$ in this limit.

But in fact similarity between $\rho(E)$ and $P(E)$ is not restricted
by the limit of strong Coulomb interaction $v_{pl}\gg v_F$. One can
obtain the full expression for the plasmon velocity, Eq. 
(\ref{specC}), by taking into account the compressibility
of the neutral Fermi gas in the derivation of the plasmon  mode.
Namely, equation of electron motion should be:
\begin{equation}
{\partial I\over \partial t}={e^2 n_1\over m} E_z -c_s^2 \nabla q~.
    \label{NwtC}\end{equation}
In the 3D case this yields the spectrum of the 3D plasma waves with the
dispersion: $\omega^2= \omega^2_{pl} + c_s^2 k^2$. Here $c_s$ is the sound
velocity, $q$ is the charge per unit wire length, and $E_z=\nabla \phi$ is
the electric field along the wire. For the 1D Fermi gas $c_s=v_F$,
$\phi=2q\ln (r_g/r_0)$, and  Eq. (\ref{NwtC}) together with the continuity
equation
\begin{equation}
{\partial q\over \partial t}+ \nabla I=0
    \label{cont}\end{equation}
yield  the sound-like spectrum with the velocity given by  Eq.
(\ref{specC}). In order to take into account the compressibility of the
neutral Fermi gas, the voltage $V$ across the junction in the Josephson
relation Eq. (\ref{Jos}) should be defined as a difference of the
electrochemical potential across the junction:
\begin{equation}
V =\phi_1-\phi_2 + {\mu_1-\mu_2\over e}~. 
      \end{equation}
Earlier the EQF theory \cite{IN} took into account only the electric
potential difference $\phi_1-\phi_2$, but neglected the chemical
potential difference $\mu_1-\mu_2$. This is usually well justified for
3D metals, but not for nanotubes. The effect of the neutral-gas 
compressibility on the plasmon velocity
and the impedance can be incorporated by introducing a renormalized
capacitance of the line,
\begin{equation}
{1\over \tilde c}={1 \over c_l} +{1 \over c_0}~,~~{1 \over c_0} ={1\over
e^2}{\partial \mu \over \partial n}={R_K v_F\over 2}~.
        \label{tr-c}\end{equation}
While the geometric capacitance $c_l$ is related to the energy of the
electric field between the wire and the  metallic ground, the capacitance
$c_0$ is related to the kinetic energy of the electron Fermi sea. Then the
impedance for an infinite transmission line formed by a 1D wire is:
\begin{equation}
 Z = { V \over I}=\sqrt{l_l \over \tilde c}=l_l v_{pl} =\sqrt{{
l_l \over c_l} +{R_K^2\over 4}} ={R_K\over 2}{v_{pl} \over v_F}~.
      \label{trans-c}\end{equation}
Finally we obtain that the probability $P(E)$ at $E< \hbar/\tau$ is
described by the same power law as the DOS $\rho(E)$ in the LL theory:
$\alpha_E=\alpha_L$. Thus both the EQF and the LL theory predict a
suppression of conductance, but physical reasons for it look different.
In the LL picture the current is suppressed because there  are no single
electron quasiparticles, and the charge is transported  by bosonic modes
(plasmons). In the Coulomb blockaded normal junction between 3D wires
single-electron states are available, but at low voltage bias  the
tunneling electron has not enough energy to get into them.

Equation (\ref{trans-c}) shows that the impedance of a wire has a quantum
limit like the d.c. resistance of a ballistic conductor. But the ``quantum
impedance'' $R_K/2$ is two times smaller than the quantum resistance $R_K$
found by Landauer for one channel (spinless electrons) \cite{Dat}. This is
because the d.c. energy dissipation is related to {\em two} contacts,
whereas the quantum impedance presents losses at {\em one} contact. Another
difference is that d.c. energy dissipation takes place in massive
electrodes, which supply a current to a ballistic conductor \cite{Dat}. But
a.c. energy dissipation given by the quantum impedance occurs inside the
wire. Indeed, in order to be considered as an infinite transmission line
(see discussion in the end of Sec. \ref{env}), the wire must have a small
but finite ohmic resistance, which is able to suppress reflection of
plasmons from another end of the wire (plasmon resonances).

Since $P(E)$ is connected with the second derivative of the $IV$ curve
and $\rho(E)$ is connected with the first derivative, the exponents for
the conductance, $\beta_E=\alpha_E$ and $\beta_L=\alpha_L-1$, differ by
unity. In the limit of the weak Coulomb interaction inside
the conductor ($\alpha_E=\alpha_L=v_{pl}/ v_F \approx 1$) the difference is
most pronounced: the LL theory predicts an ohmic $IV$ curve, whereas the EQF
theory predicts the Coulomb blockade regime with the power law $I \propto
V^2$, if $V \ll e/C_T$. This so-called ``weak-Coulomb-interaction'' limit
is valid at the condition that $v_F \gg \ln(r_g/r_0)/R_K$.

\section{Tunneling into a multichannel quasi-1D conductor} \label{Lut-2}

One may expect that the difference between the EQF and the LL
predictions should vanish if a number of channels (electron modes)
grows. This was confirmed by the work of Matveev and Glazman\cite{MG}.
But some features of this crossover remain unclear. Matveev and
Glazman\cite{MG} have shown that in the limit of a large number of
channels both theories predict the same power law for conductance:
$\beta_E=\beta_L$. But this means that the exponents $\alpha_E$ and
$\alpha_L$ for $P(E)$ in the EQF theory and for $\rho(E)$ in the LL theory
differ by 1, in contrast to the one-channel case, in which the
exponents for $\rho(E)$ and for
$P(E)$ coincide, but the exponents for conductance are different. On the
other hand, due to reasons explained in the previous section, $\rho(E)$
and
$P(E)$ should have the same functional dependence on $E$ at small
energies.

For further discussion of this problem, I rederive the result of Matveev
and Glazman in terms of the transmission-line modes. A quantum
wire with $N$ electron channels can be considered as a collection of $N$
parallel transmission lines. We assume that the electrostatic energy in
this circuit is
\begin{equation}
E_{el}= \sum _{i=1}^N {q_i^2\over 2c_i}  +{1\over 2c_l}\left(\sum _{i=1}^N
q_i\right)^2~,
      \end{equation}
where $c_l=1/(2 \ln{r_g/r_0})$ is the capacitance per unit length
between each channel and the ground, and the capacitances $c_i$ are
related with the electron kinetic energies, which can be
different for various channels in general. The voltages, which are
defined as electrochemical potentials with respect to the ground, are: 
\begin{equation}
V_i={\partial E_{el}\over \partial q_i}={q_i\over c_i} +
{1\over c_l} \sum _{j=1}^N q_j~.
           \end{equation}
Using the continuity equations,
\begin{equation}
{\partial q_i\over \partial t}+{\partial I_i\over \partial x}=0,
           \end{equation}
and the equations of motion for electrons in all channels,
\begin{equation}
 l_i {\partial I_i \over \partial t}=-{\partial V_i\over \partial x}~,
    \label{e-m} \end{equation} 
we obtain $N$ equations for charge densities $q_i$. Here $l_i$ is the
inductance per unit length of the $i$th channel. For the plane wave
solutions $q_i\propto e^{ikx-i\omega t}$, these $N$ equations are:
\begin{equation}
s^2 l_i q_i={q_i\over  c_i}+ {1\over c_l} \sum _{j=1}^N q_j~,
       \label{q-s}       \end{equation} 
where $s=\omega/k$ is the velocity of the sound-like modes. Equations
(\ref{q-s}) are identical to Eqs. (11) of Matveev and Glazman \cite{MG}.
In terms of parameters, used by  Matveev and Glazman, the interaction
potential $V$ and the Fermi velocity $v_i$ of the $i$th channel, the
inductance and the Fermi velocity of the $i$th channel are $l_i=c_lV/v_i$
and $v_i=1\sqrt{l_ic_i}$.

One must find $N$ eigenmodes of the system. Further we assume that all
channels are identical: $c_i=c_0$, $l_i=l_l$. Then there are only two
eigenvalues:  
\begin{equation}
s_1^2= {1\over l_l c_0}+ {N \over l_l c_l}
              \end{equation} 
and $s_2^2 =1/l_l c_0$, which is $(N-1)$-degenerate. The first eigenvector
is
$q_i^{(1)}\propto 1$. This is a plasmon mode, in which the total charge
$\sum _{j=1}^N q_j$ oscillates and the velocity $s_1$ is the plasmon
velocity $v_{pl}$. The other
$N-1$ modes are neutral, and therefore their velocity $s_2$ coincides with
the Fermi velocity
$v_F=1/\sqrt{l_l c_0}$. The $N-1$ eigenvectors for the eigenvalue $s_2^2$
can be chosen as ($j
\neq 1$)
\begin{equation}
q_i^{(j)} \propto \delta_{ij} - {1\over N}~.
              \end{equation} 

Now we want to calculate the impedance. In the $N$-channel system the
impedance is a matrix: $Z_{ij}= V_{i}/I_j$. In order to find it, one
must calculate voltages $V_i$ of various channels assuming that the current
is present only in the $j$th channel. The densities of state for various
channels, which are considered in the LL model, are associated with the
diagonal elements of the impedance matrix. In our case all channels are
equivalent, and one can calculate the impedance $Z_{11}$ for the first
channel. One should find a superposition of $N$ eigenvectors, which
determines the charge
\begin{equation}
q_i= \sum _{j=1}^N A_j q_i^{(j)}=A_1 + A_i (1-\delta_{1i})-{1\over N}\sum _{j=2}^N
A_j~, 
              \end{equation} 
the current 
\begin{equation}
I_i= \sum _{j=1}^N A_j s_j q_i^{(j)} ~,
              \end{equation}
and the voltage 
\begin{equation}
V_i= \sum _{j=1}^N A_j l_l s_j^2 q_i^{(j)} 
              \end{equation} 
in any channel. The coefficients $A_j$, which define the superposition,
should be found  from the conditions that the total current comes only
into the 1st channel. The condition for the 1st channel is:
\begin{equation}
I = A_1 s_1 - \sum _{j=2}^N {A_j s_2\over N }~.
              \end{equation} 
For the other $N-1$ channels ($i\neq 1$):
\begin{equation}
0 = A_1 s_1+A_i s_2 - \sum _{j=2}^N {A_j s_2\over N}~.
              \end{equation} 
This system of equations has a solution: $A_1= I/Ns_1$ and for $i\neq 1$ 
$A_i=-I/s_2$. Now the voltage at the first channel is:
\begin{equation}
V_1  = l_lI \left(A_1 s_1^2 - \sum _{j=2}^N {A_j s_2^2\over N
}\right) =  l_l I \left({1\over N}s_1 + {N-1 \over N }s_2\right)~.
              \end{equation}
Finally the impedance is 
\begin{equation}
Z_{11}={V_1 \over I} ={1\over N}\sqrt{{l_l\over c_0}+{N l_l\over c_l}}+
{N-1
\over N}\sqrt{l_l\over c_0}=\sqrt{{R_K^2\over 4N^2 }+ {l_l\over N c_l}}
+{N-1\over N}{R_K\over 2}~. 
              \end{equation}
The ``electric'' part of this impedance, $\sqrt{l_l/Nc_l}$, is related to
a simple fact that the inductance of $N$ parallel transmission lines is
in $N$ times less than the inductance $l_l$ of one transmission line. 
 
In terms of the impedance $Z_{11}$ the exponent $\alpha_L$ is
\begin{equation}
\alpha_L= {2Z_{11}\over R_K}= {1 \over N}{v_{pl} \over v_F}+{N-1\over N}~.
     \label{alL}         \end{equation}
One can compare it with the EQF exponent for a multichannel system:
\begin{equation}
\alpha_E= {2Z\over R_K}= {1 \over N}{v_{pl} \over v_F}~.
      \label{alE}        \end{equation}
This directly follows from the one-channel value, bearing in mind that
the multichannel impedance is an impedance of $N$ parallel transmission
lines. Thus the exponents $\alpha_L$ and $\alpha_E$ are connected with
the impedances by the same relations, which follow from the
fluctuation-dissipation theorem, but the impedances are different in the
EQF and the LL approaches. The EQF theory ignores the neutral modes,
which are important in the LL approach. In order to obtain the EQF
impedance one should assume that the current
$I$ is distributed uniformly between channels, and therefore the neutral
modes are not excited. On the other hand, the EQF approach takes into
account the single-particle excitations, which are absent in the LL
approach, but the neutral modes play the same role, and eventually the
exponents for conductance are the same for multichannel conductors.

The LL and EQF conductance exponents, which follow from Eqs. (\ref{alL})
and  (\ref{alE}), are very close each other for a large number $N$ of
channels: 
\begin{equation}
\beta_L=\alpha_L-1= {1 \over N}\left({v_{pl} \over v_F}-1\right)~,~~~
\beta_E=\alpha_E= {1 \over N}{v_{pl} \over v_F}~.
     \label{con}         \end{equation}

However, the EQF theory predicts the conductance exponent, which is similar
to that in the LL theory, only in the Coulomb blockade regime $V \ll e/C_T$.
At higher voltages the junction capacitance $C_T$ shunts the environment impedance, and the $IV$ curve is given by the asymptotic expression Eq. (\ref{IVh}). This corresponds to constant conductance:
$\beta_E=\alpha_E=0$. If in analogy with the EQF theory one tries to
introduce the capacitance $C_T$ into the effective circuit in the LL
picture, this yields a dramatically different result. The capacitance $C_T$
short circuits the impedance $Z_{11}$, like in the EQF picture, and
$\alpha_L \rightarrow 0$. But now the conductance exponent $\beta_L
\rightarrow -1$, and the LL theory predicts the voltage-independent current
at high voltages. This contradicts to experiment \cite{Penttila00}, and it
is unclear what is a proper way to take into account the junction Coulomb
energy in the LL theory.

\section{Discussion and summary} \label{Con}

The analysis has shown that despite a conceptual difference between the EQF
and the LL approaches, their predictions for the power law exponents of the
conductance are the same for multichannel conductors. Therefore one
cannot discriminate between two approaches in studying the power law
exponents for a quasi-1D conductor with a large number of conduction
channels.  

This means that it is difficult to detect the LL behavior in multiwall
nanotubes. For single-wall nanotubes the difference between the LL and EQF
predictions is more pronounced, but also not very large. Every nanotube
wall has four channels (two due to spin, and two due to two bands). This
means that for a single-wall nanotube the difference between the
conductance exponents $\beta_E$ and $\beta_L$  is only 0.25 for the same
ratio $v_{pl}/v_F$. In order to reveal this difference and thus to
discriminate between two pictures, it would be useful to measure not only
the $IV$ curve, but also the plasmon velocity for the same nanotube using
contactless methods.

 Luttinger liquid is a well established and irrefutable concept of
the condensed-matter physics, which has been confirmed by many years of
theoretical studies. However, its experimental verification remains to be
a challenging problem, which requires further efforts both in experiment
and theory.

\section*{ACKNOWLEDGMENTS}

This work is a result of collaboration with the Nanogroup of Low
Temperature Laboratory of Helsinki University of Technology. I thank Pertti
Hakonen and Mikko Paalanen for a stimulating interest to this work and
numereous discussions. I also appreciate very much discussions of results
with Boris Altshuler, David Haviland, and Nils Schopohl during the present
symposium on Ultra Low Energy Physics. The work was supported by the Academy
of Finland, by the Israel Academy of  Sciences and Humanities, and by the
Large Scale Installation Program ULTI-3 of the European Union.

\end{document}